\documentclass[graybox]{svmult}

\usepackage{mathptmx}       
\usepackage{helvet}         
\usepackage{courier}        
\usepackage{type1cm}        
\usepackage{makeidx}         
\usepackage{graphicx}        
\usepackage{multicol}        
\usepackage[bottom]{footmisc}
\usepackage{epsfig}

\makeindex             

\begin{document}

\title*{The ANTARES Neutrino Telescope}
\author{Giorgio Giacomelli, for the Antares Collaboration
\\ University of Bologna and INFN Bologna
\\  \email{giacomelli@bo.infn.it}}

\maketitle

\vspace{-3cm}

{\bf Invited talk at the Conference on "New Trends in High-Energy Physics $\&$ Safe Nuclear Energy"}
\\
Yalta, September 2008
\vspace{1.5cm}

\abstract{  The ANTARES underwater neutrino telescope, at a depth of 2475 m in the Mediterranean Sea, near Toulon, is taking data in its final configuration of 12 detection lines. Each line is equipped with 75 photomultipliers (PMT) housed in glass pressure spheres arranged in 25 triplets at depths between 100 and 450 m above the sea floor. The PMTs look down at 45$^{\circ}$ to have better sensitivity to the Cherenkov light from upgoing muons produced in the interactions of high energy neutrinos traversing the Earth. Such neutrinos may arrive from a variety of astrophysical sources, though the majority are atmospheric neutrinos. The data from 5 lines in operation in 2007 yielded a sufficient number of downgoing muons with which to study the detector performances, the vertical muon intensity and reconstruct the first upgoing neutrino induced muons.   }


\section{Introduction}\label{sec:1}
The effort to build large sea water Cherenkov detectors was pioneered by the Dumand Collaboration with a prototype at great depths close to the Hawaii islands [1]; the project was eventually cancelled. Then followed the fresh water lake Baikal detector at shallow depths, which was later enlarged [2]. Considerable progress was made by the Amanda and Icecube ice telescopes in Antarctica [4, 5]. In the Mediterranean sea the NESTOR collaboration tested a deep line close to the Greek coast [5] and the NEMO Collaboration close to Sicily [6];  the ANTARES neutrino telescope at 2475 m below sea level was deployed in the Mediterranean sea, close to Toulon, France, see Figs. 1, 2, 3 [7, 8].

%
\begin{figure}[t]
\begin{center}
\vspace{0.5cm}
\sidecaption
\includegraphics[scale=.28]{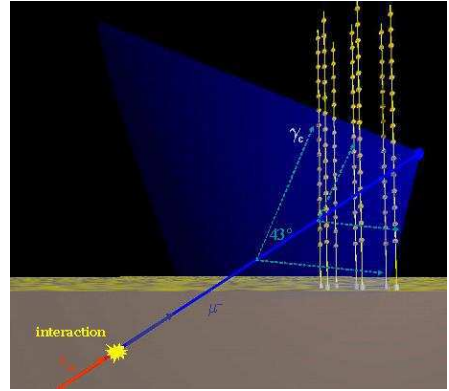}
%
\caption{ Optical detection of $\nu_{\mu}$ neutrinos interacting in the rock below the sea and giving an upgoing muon which yields Cherenkov radiation in the detector.}
\label{fig:1}       
\end{center}
\end{figure}

The completed ANTARES neutrino telescope, at 2475 m below sea level, is presently taking data with 12 detection lines, held vertical by buoys and anchored at the sea floor. The lines are connected to the Junction Box (JB) that distributes power and data from/to shore. The instrumented part of each line starts at 100 m above the sea floor, so that Cherenkov light can be seen also from upgoing muons coming from neutrino interactions in the rock below the sea, Fig. 1. The lines are separated by $\sim$70 m, and are on an approximate octagonal structure. Each line has 25 floors (storeys), each with 3 Hamamatsu 10'' Photomultipliers (PMT) looking downward at 45$^{\circ}$ from the vertical, housed inside pressure resistant glass spheres (Optical Module (OM)), Fig. 3. The storeys include frontend electronics for signal processing and digitization [9, 10, 11, 12]. 

%
\begin{figure}[b]
\begin{center}
\sidecaption
\includegraphics[scale=.49]{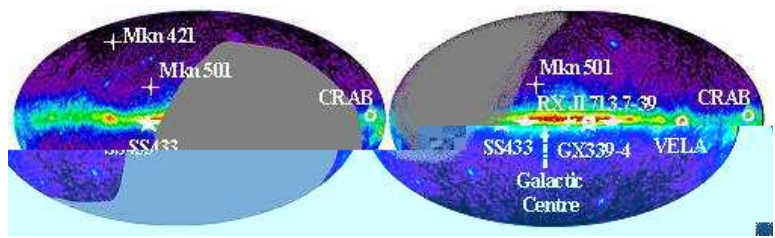}
%
\caption{ Left: Regions of sky observable by neutrino telescopes at the South Pole (Amanda, IceCube); Right: in the Mediterranean sea (Antares at 43$^\circ$ North).}
\label{fig:2}       
\end{center}
\end{figure}

One of the main reasons to build neutrino telescopes is to study high energy muon neutrino astronomy. Neutrinos may be produced in far away sources, where charged pions are produced and decay, like very high energy photons from $\pi^0$ decay, and reach the earth undisturbed. Instead the photons interact with the Cosmic Microwave Background (CMB) radiation and with matter, protons are deflected by magnetic fields and neutrons are unstable. The main drawback of neutrinos is that one needs very large detectors. 
Neutrino telescopes may also allow dark matter searches [13], searches for exotic particles (fast magnetic monopoles, nuclearites, etc. [14, 15, 16, 17], may contribute to the study of atmospheric neutrino oscillations [18]) and test conservation laws [19].

It is worthwhile to recall that neutrino telescopes in the Northern hemisphere have access, can see, the center of our galaxy, while neutrino telescopes in Antarctica cannot see it, Fig. 2. Thus a large telescope in the Mediterranean sea (KM3) [20], and one in Antarctica (Icecube) are complementary and scientifically justified.

\section{The ANTARES experiment}\label{sec:2}

%
\begin{figure}[t]
\begin{center}
\vspace{0.2cm}
\sidecaption
\includegraphics[scale=.35]{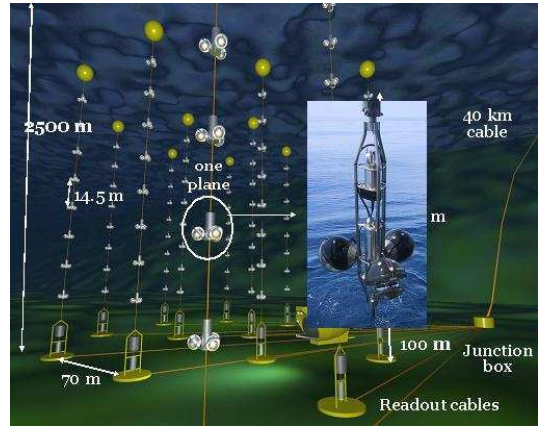}
%
\caption{ Scheme of the Antares completed detector.}
\label{fig:3}       
\end{center}
\end{figure}

Fig. 3 shows an artistic illustration of the 12 lines Antares detector; Fig. 4 shows the detector as seen by downgoing muons. The main features of ANTARES were recalled in the Introduction; here we shall recall some other features.

The water properties at the site were extensively studied and are given in ref. [7]: the absorption length in blue light is $\sim$60 m, the effective scattering length is $\sim$250 m. Several prototype lines and an instrumentation line were deployed in 2003-2005 [9, 10]. The final first 2 lines were deployed in 2006; in early 2008 the telescope was completed. The electro-optical cable to shore was repaired in mid 2008.

%
\begin{figure}[t]
\begin{center}
\sidecaption
\includegraphics[scale=.43]{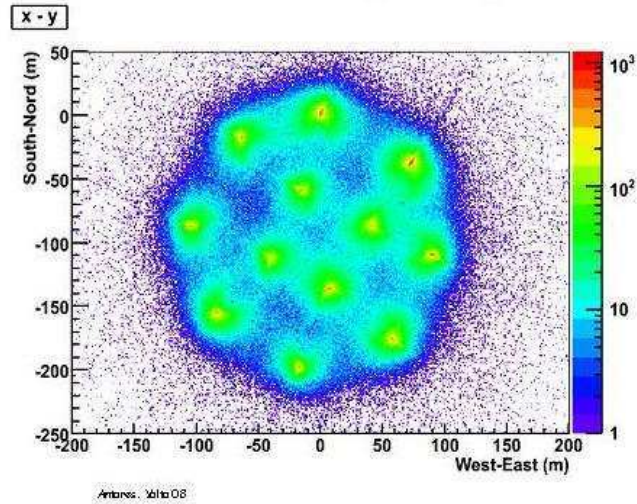}
%
\caption{ The 12 line Antares detector as seen by downgoing atmospheric muons.}
\label{fig:4}       
\end{center}
\end{figure}

Large photomultipliers, optical modules and the data acquisition system (DAQ) were extensively studied [8, 9, 13]. Each OM has a mu-metal schield for the Earth magnetic field, and contains a system of LED and laser optical beacons for calibrations and relative calibrations of different OMs. Precision timing is distributed via a 200 MHz clock sygnal with a precision of $\sim$100 ns. The final timing precision is $\sim$0.5 ns, which yields an angular resolution of 0.3$^\circ$ for muons with energies $>$10 TeV. The readout system assigns time-stamps to PMT signals, digitizes their charges and merges the data from the 75 PMTs on each detection line in a single fiberoptic [13]. The data which satisfy a minimal requirement, L0 hits, above a threshold of 1/3 photoelectron, are sent to shore. The reconstruction of muon tracks is based on ns measurements of the arrival times of Cherenkov photons at the PMTs. This requires precise knowledge of the relative positioning of OMs. An array of acoustic trasponders is deployed on the sea floor; they transmit sound signals to hydrophones mounted at five different altitudes on each detection line. Several special triggers are available ($\sim$7.5 Gb/s). A filter removes excessive noise from bioluminescence and $\beta$ decays from K$^{40}$ decays. Usually this noise is at the level of $\sim$60 kHz, but there was a higher rate in the first data taking periods, in 2006 and part of 2007, characterized by large sea currents.

For the 12 line detector, Figures 5 and 6 show the on line event display examples of one muon bundle and of one neutrino candidate event, respectively. The present rate of neutrino induced upgoing muons is $\sim$5 events/day.

\section{Preliminary results}\label{sec:3}

The first Antares detector line was deployed in February 2006 and took data alone till October 2006. During this period the background rates were often high and one had to select data over a more limited time [21] and several problems were solved, like the reconstruction software. The downgoing muon data were used to calculate the vertical muon intensity versus depth, Fig. 7: Note the good agreement with previously measured data which were taken with different types of detectors.

%
\begin{figure}[t]
\begin{center}
\sidecaption
\includegraphics[scale=.43]{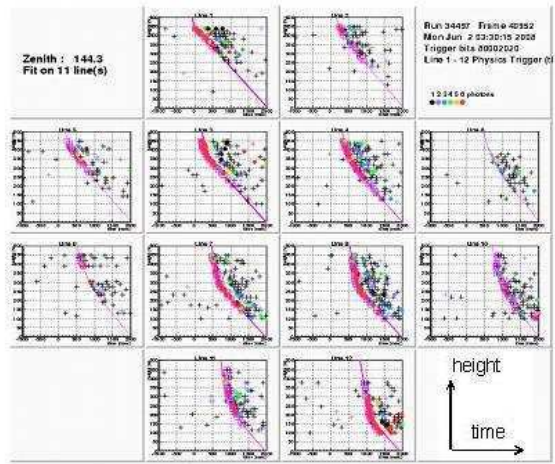}
%
\caption{ Example of a muon bundle seen by the 12 line Antares telescope.}
\label{fig:5}       
\end{center}
\end{figure}

%
\begin{figure}[b]
\begin{center}
\hspace{0.7cm}
\sidecaption
\includegraphics[scale=.53]{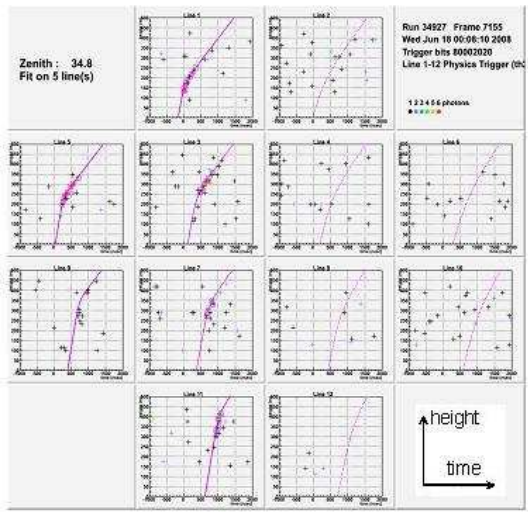}
%
\caption{ Example of a neutrino candidate seen by the 12 line Antares telescope.}
\label{fig:6}       
\end{center}
\end{figure}

Thanks to its modular project Antares was able to take data in a non complete configuration: it took data with a 5 line configuration in the period February-November 2007. This period of data taking was adequate to obtain a good sample of downgoing atmospheric muons, to improve the track recontruction software with several lines, to improve and test several Monte Carlo codes [31, 32, 33, 34].   

Figs 8-9 show the zenith and azimuth distributions of reconstructed downgoing and upgoing muon tracks. A loose quality cut was applied, that is not able to remove all the badly reconstructed tracks, which are still dominant in the upgoing track region (0$^\circ$$<$$\vartheta$$<$90$^\circ$). The black points are the data; the solid line refers to MC expectations obtained using the full CORSIKA simulation with QGSJET01 for the hadronic interaction description [32], and the Horandel model [33] of primary cosmic rays: a good agreement is evident both in shape and in absolute normalization. The shadowed band is an estimate of the systematic effects due to the uncertainties on environmental and geometrical input parameters in the Monte Carlo simulation. In Fig. 8, the dotted line is obtained with CORSIKA{ QGSJET01 and the NSU model of primary cosmic rays [32]. The difference between the two MC expectations can be ascribed to the different composition models. The dashed-dotted line refers to the fast parametric simulation using MUPAGE [34]. This simulation is based on an all-particle CR flux obtained in underground experiments. The difference with respect to the predictions obtained with the NSU model are likely due to the different hadronic interaction description (DPMJET). Globally, data and MC show a good agreement in shape and are in agreement within 30 - 40$\%$ in absolute normalization.

%
\begin{figure}[b]
\begin{center}
\sidecaption
\includegraphics[scale=.33]{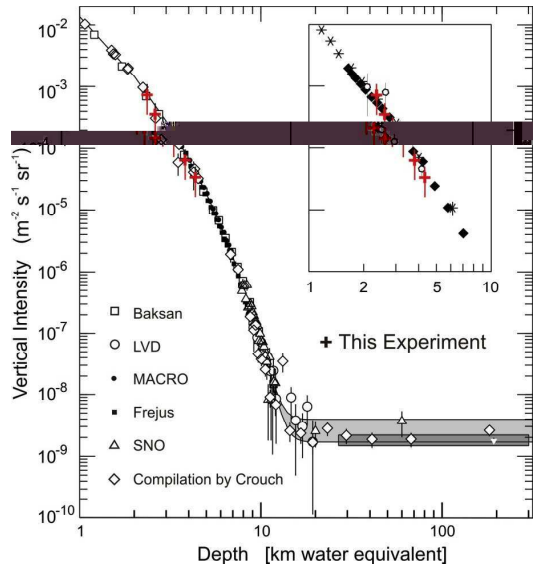}
%
\caption{ Vertical muon intensity versus depth. Crosses are from the first Antares line [21]; other data are from: the compilations of Crouch [22], Baksan [23], LVD [24], MACRO [25], Frejus [26], and SNO [27]. The shaded area at large depths represents neutrino induced muons of energy above 2 GeV. In the insert the Antares data are compared to other water or ice data [28, 29, 30]. }
\label{fig:7}       
\end{center}
\end{figure}

In Fig. 9 some peaks are visible in the azimuth data distribution, which are well reproduced by the MC simulation (CORSIKA-QGSJET-Horandel model). They correspond to the enhancement of the muon reconstruction efficiency for tracks lying on a plane defined by two or more ANTARES strings. The good superposition between data and MC expectations indicates that the positioning systems are working well and that MCs can reproduce the main geometrical features of the detector.

%
\begin{figure}[t]
\begin{center}
\vspace{0.3cm}
\sidecaption
\includegraphics[scale=.47]{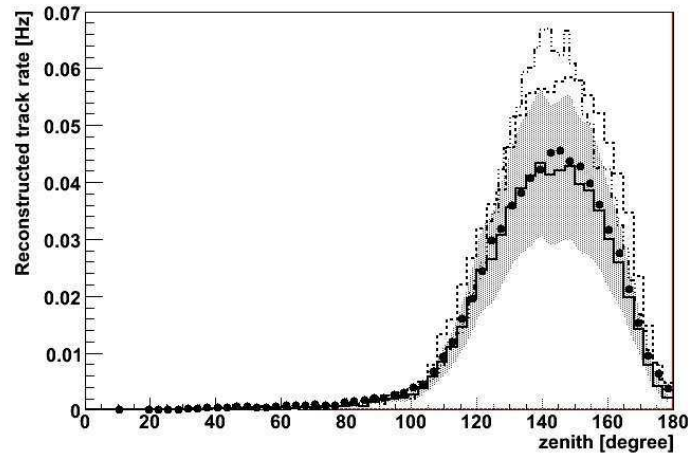}
%
\caption{ (Preliminary) Zenith distribution of reconstructed tracks with 5 Antares lines. Black points are data, lines refer to MC predictions: the solid line histogram is the full simulation with CORSIKA, QGSJET01 and the Horandel model for CR composition; the dotted line is the same MC reweighed for the NSU CR model. The dashed-dotted line is the parametric simulation with MUPAGE. The shadowed band is the systematic uncertaintys due to environmental and geometrical parameters.}
\label{fig:8}       
\end{center}
\end{figure}

\section{Conclusions}\label{sec:4}

The Antares neutrino telescope is running in its final configuration of 12 detection lines. The preliminary results obtained with a single line and with 5 lines allowed checking and improving the reconstruction programs, and testing of MC codes. 

The data taken with line 1 measured the vertical atmospheric muon intensity as a function of depth: the distribution agrees well with previous measurements.

The zenith and azimuth distributions of atmospheric downgoing muons measured with 5 lines agree with MC expectations, which have a $\sim$30$\%$ systematic uncertainty; the main source of uncertainty lies in the primary cosmic ray model.\\

I thank drs. S. Cecchini, A. Margiotta, M. Errico and prof. M. Spurio for their cooperation. I thank dr. M. Errico for typing and correcting the manuscript.

%
\begin{figure}[t]
\begin{center}
\sidecaption
\includegraphics[scale=.47]{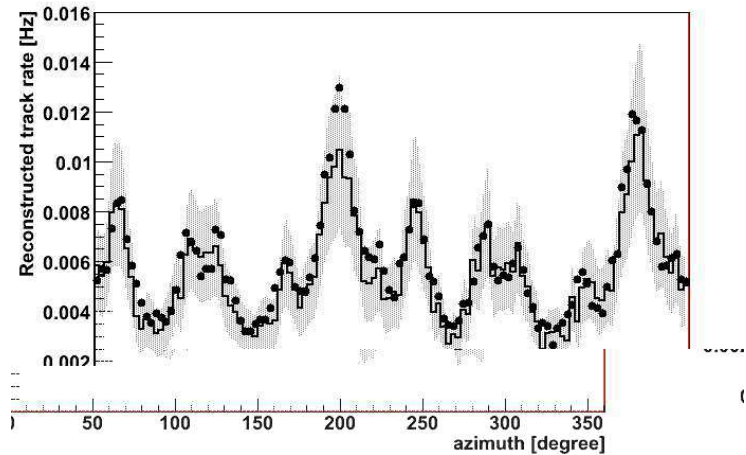}
%
\caption{  (Preliminary) Azimuth distribution of reconstructed tracks with 5 Antares lines. Black points are data. The solid line histogram is the MC expectation with the CORSIKA code + QGSJET01 for air shower simulation plus Horandel model for CR composition. The shadowed band is an estimate of the systematics uncertanty due to environmental and geometrical parameters. }
\label{fig:9}       
\end{center}
\end{figure}

\end{document}